\documentclass[twocolumn,showpacs,preprintnumbers,amsmath,amssymb]{revtex4}
\usepackage{graphicx}% Include figure files
\usepackage{dcolumn}% Align table columns on decimal point
\usepackage{bm}% bold math

\newcommand{\abs}[1]{|#1|}

\renewcommand{\i}{{\rm i}}

\newcommand{\e}{{\rm e}}

\begin{document}

\preprint{donorarray17}

%\preprint{APS/123-QED}

\title{Excited states of defect lines in silicon: A first-principles study based on hydrogen cluster analogues}
\author{W.~Wu}
\email{wei.wu@ucl.ac.uk}
\author{P. T.~Greenland}\author{A. J.~Fisher}
\affiliation{UCL Department of Physics and Astronomy and London Centre for Nanotechnology,\\
University College London, Gower Street, London WC1E 6BT}
\author{Nguyen H. Le}
\author{S.~Chick}\author{B. N.~Murdin}
\affiliation{Advanced Technology Institute and Department of Physics, University of Surrey, Guildford GU2 7XH, United Kingdom}
\date{\today}%
\begin{abstract}
Excited states of a single donor in bulk silicon have previously been studied extensively based on effective mass theory. However, a proper theoretical description of the excited states of a donor cluster is still scarce. Here we study the excitations of lines of defects within a single-valley spherical band approximation, thus mapping the problem to a scaled hydrogen atom array.  A series of detailed full configuration-interaction and time-dependent hybrid density-functional theory calculations have been performed to understand linear clusters of up to 10 donors. Our studies illustrate the generic features of their excited states, addressing the competition between formation of inter-donor ionic states and intra-donor atomic excited states. At short inter-donor distances, excited states of donor molecules are dominant, at intermediate distances ionic states play an important role, and at long distances the intra-donor excitations are predominant as expected. The calculations presented here emphasise the importance of correlations between donor electrons, and are thus complementary to other recent approaches that include effective mass anisotropy and multi-valley effects. The exchange splittings between relevant  excited states have also been estimated for a donor pair and for a three-donor arrays; the splittings are much larger than those in the ground state in the range of donor separations between 10 and 20 nm.  This establishes a solid theoretical basis for the use of excited-state exchange interactions for controllable quantum gate operations in silicon.
\end{abstract}

% PACS, the Physics and Astronomy Classification Scheme.

\maketitle
\section{Introduction}\label{sec:introduction}
After decades of development and incorporation of many new materials, the core material technology of microelectronics remains based on silicon. Impurities in silicon play a vital role in its transport, magnetic, and optical properties \cite{zwanenburg2013}. The recent encouraging progress in deterministic positioning of dopants in silicon \cite{schofield2003, brazdova2016, usman2016} promises atom-by-atom design and bottom-up fabrication of silicon-based nano-devices; such nano-structures can offer not only an ultimate limit for conventional electronic components such as wires \cite{Weber:2012gv} and tunnel structures \cite{Pascher:2016hj}, but also a potential platform for many applications in silicon quantum electronics \cite{zwanenburg2013}, and ultimately for new technologies that exploit the quantum properties of electron spin and orbital motion \cite{hollenberg2006, BMR2013, salfi2016, BGK2016, GBK2016, BWK2017}. An obvious candidate for a quantum bit (qubit) is a donor electron spin: the spin-lattice relaxation time ($T_1$) of donor electron spins in silicon has been measured to be up to a few thousand seconds \cite{morley2010, tyryshkin2012}, and the coherence time ($T_2$) is up to ms, limited only by interactions with neighbouring electron or nuclear spins. The $T_2$ can be enhanced further, to several seconds by the use of field-insensitive 'clock transitions' \cite{wolfovicz2013}, and even further by the use of isotopically pure $^{28}$Si.

Recent progress has shown that the orbital degree of freedom of a dopant electron in silicon can also be controlled and could potentially itself serve as a qubit. Terahertz (THz) optical excitations (tuned to an energy-level spacing of $\sim$meV) can be used to manipulate and detect Rydberg states of donors by using a free-electron laser \cite{cole2001, thornton2008, litvinenko2015, chick2017}. In the density range where donor pairs are dominant, the optical field has been used to detect and control the electron tunnelling between donor pairs of phosphorus and antimony \cite{litvinenko2014}, while in three-donor clusters optical excitation (de-excitation) of a shallower 'control' donor has the potential to switch on (off) the exchange interaction between the other two deeper donors, thus forming an optically controlled quantum gate \cite{sfg, wu2007}. %Moreover, the analogy between shallow impurities in silicon and hydrogen atoms has led to renewed interest in the high-field mageto-optics of hydrogens in neutron stars \cite{murdin2012}. 
There has also been growing experimental interest in performing quantum simulations \cite{salfi2016, HFJ2017} in donor clusters.  Arrays of dopants in silicon \cite{salfi2016} and quantum dots \cite{HFJ2017} have been fabricated for the quantum simulation of the Fermi-Hubbard model; the freedom to position the atoms arbitrarily enables tuning of correlations by varying the inter-donor distance \cite{nguyen2017}.  Such a platform could be enhanced by the ability to probe the state of the electrons using optical absorption.  It is therefore timely to study theoretically the optical properties of multi-atom donor clusters such as arrays, and in particular the effect of excitation on the spin-spin interaction. 

%In addition, experiments on impurity clusters in silicon as a platform for quantum information processing have gained a significant surge recently [BMR2013, SMR2016, BGK2016, GBK2016, BWK2017]. 
The electronic structure of a single dopant in silicon (or germanium) has been extensively studied previously \cite{luttinger1954, kohn1955, foulkner1969, PWU2014}. There are mainly two types of methodologies, including effective mass theory (EMT) and atomistic tight-binding (ATB) methods. Within EMT, either anisotropic hydrogenic trial wave functions \cite{kohn1955} or a Coulomb potential deformed through a coordinate transformation \cite{foulkner1969} have been used to account for the anisotropy of the conduction-band minimum. These calculations were refined in the 1970s by adding multi-valley effect (MVE) and the effects of deviations from a pure Coulomb potential \cite{ning1971}, producing agreement with experimental spectra. Recently Gamble, et. al. solved the Shindo-Nara multi-valley equation \cite{sn1976} by including the full Bloch wavefunctions of silicon \cite{gamble2015}, showing a good agreement with the experimental energy spectrum of single phosphorus atoms, and also gave theoretical values for donor-donor tunnel couplings. The ATB calculation is another commonly used method, which considers the full lattice structure of the host material \cite{RWB2007, LRW2008}. Both approaches have produced theoretical results in excellent agreement with the experimental ground-state energy spectrum and hyperfine Stark shift of single phosphorus donors in silicon. For two electrons in a single donor ($D^-$), ATB calculations followed by a self-consistent Hartree method were used to account for the electron-electron interaction in a mean-field way, while neglecting exchange \cite{HBT2014, WCK2016}. A full configuration-interaction (FCI) computation has been performed for $D^-$ with single-electron wavefunctions obtained from the atomistic tight-binding method \cite{WTH2016, TSB2017}. 

%On the theory side, clusters of dopants in silicon was studied with effective mass theory [24-27] and its extension [WH2005,28]. In these works, the interaction between the excess electrons is either treated with the Heitler-London formalism [27,28, WH2005] or neglected [24-26]. 
Studies of impurity clusters in silicon have gained a significant surge recently \cite{dusko2016, saraiva2015, klymenko2017}. The exchange interactions between donors have been studied within EMT, combined with the Heitler-London formalism \cite{koiller2002}, including variable binding energies \cite{wu2008}. The electronic structure of a donor pair has been studied by using CI within the $1s$ manifold \cite{saraiva2015}. Larger donor clusters have been studied by using density-functional theory (DFT) and the GW method (for bulk silicon) combined with EMT \cite{klymenko2017}; this approach included implicitly the electron correlations in the bulk, but missed the explicit Coulomb interaction between electrons within the multi-donor 'molecule'.  A combination of EMT and ATB has been employed to calculate the electronic structure of thin dopant chains and to study the localisation of donor electrons owing to disorder \cite{dusko2016}. In the calculations to date the electron correlations are either at least partly absent, or confined to the lowest manifold of the 1s ground states of single donors and donor pairs.

The multi-valley effect and central-cell corrections (CCC) are important for the $1s$ ground states where the electron is close to the defect core, but not so important for the more diffuse excited states. On the other hand, in the description of the electronic structure of donor clusters, the electron-electron correlations are known to play a vital role in both optical and transport properties \cite{thomas1981}. For example, the optical absorption shows a strong signature of the ionic state of a donor pair ($D^+$-$D^-$ state, also called a charge-transfer state in Ref.\cite{thomas1981}), in which an electron hops from one donor to the other, leaving a hole behind. The ionic state here is effectively a bound state of the holons and doublons that are used to analyze excitations of the Hubbard model in solid-state physics \cite{jeckelmann1999, ah2008}.  However, such low-energy ionic states appear only if proper account is taken of the intra-cluster correlations. In addition, the spherical-band approximation (replacing the anisotropic effective mass by using a single average one) turned out to be good in predicting the ground-state energy of donors in silicon \cite{ning1971}. Taken together, these facts suggest that a combination of an isotropic Hamiltonian, and wave-function within the spherical band approximation \cite{ning1971}, with highly accurate first-principles methods to treat electron correlation is a suitable starting point to describe the excited states of donor clusters. 

Here we report a systematic study of the orbital excited states and related exchange splittings of donor arrays in silicon, within the isotropic approximation to effective-mass theory but retaining a full treatment of correlations among the donor electrons. In our calculations, we use hydrogen atoms to represent silicon donors, then compute the excited states by using CI and time-dependent density-functional theory (TDDFT), and at the end scale the excitation energies by using the effective mass and dielectric constant of silicon. From these calculations, we are able to obtain a rich spectrum of physics for the excited states. We have performed FCI calculations for linear arrays consisting of up to three donors, in which the electron correlations are fully taken into account, which we used to benchmark the exchange-correlation functional in TDDFT.  Our TDDFT calculations provide a good approximation to the CI results for these small arrays, and then are extended to describe the excited states of arrays consisting of up to 10 donors. From the perspective of molecular physics, the electronic structure of H$_2$ is very well known, but the solid-state environment fixes the donor separations at implantation; hence these calculations emphasise molecular excited states in unstable hydrogen clusters far from equilibrium.  

The rest of the paper is organised as follows: we introduce the computational details in \S\ref{sec: computationaldetails}, discuss both CI and TDDFT results in \S\ref{sec: results}, and draw some general conclusions in \S\ref{sec: conclusion}.

%\begin{equation}\label{eq:multibandeffmass}
%\psi(\vec{r})=\sum_n \alpha_nF_n(\vec{r})\phi_{n\vec{k}_0}(\vec{r}),
%\end{equation}

%\begin{eqnarray}\label{eqn:a0toa5}
%\alpha^{(0)}&=&\frac{1}{\sqrt{6}}(1,1,1,1,1,1).
%\\\alpha_1&=&\frac{1}{\sqrt{12}}(1,1,1,1,-2,-2)
%\\\alpha_2&=&\frac{1}{2}(1,1,-1,-1,0,0)
%\\\alpha_3&=&\frac{1}{\sqrt{2}}(1,-1,0,0,0,0)
%\\\alpha_4&=&\frac{1}{\sqrt{2}}(0,0,1,-1,0,0)
%\\\alpha_5&=&\frac{1}{\sqrt{2}}(0,0,0,0,1,-1)
%\end{eqnarray}
\section{Computational details}\label{sec: computationaldetails}
We work in the single-valley isotropic approximation to effective mass theory, in which a shallow donor is a direct analogue of a hydrogen atom.  We therefore neglect (i)  the anisotropy of the conduction band, (ii) deviations of the potential from Coulomb form (in particular 'central-cell' corrections) and (iii) any resulting inter-valley coupling.  However we include a careful treatment of the correlations between the bound electrons.
\subsection{Effective-mass theory}
Within the single-valley approximation, the effective-mass equation \cite{luttinger1954, kohn1955} reads:
\begin{eqnarray}
[\epsilon_n(\vec{k}_0+\frac{1}{i}\nabla)+U]F_n&=&\epsilon F_n,
\end{eqnarray}
where the band energy $\epsilon_n$ is expanded around the band extremum $\vec{k}_0$ to second order-terms in $(1/i)\nabla$.  $F_n$ is the envelope function, in terms of which the  donor wavefunction is expanded using

\begin{eqnarray}\label{eq:effectivemasssum}
\psi=\sum_n \alpha_n F_n(\vec{r})\phi_{n\vec{k}_0}(\vec{r}),
\end{eqnarray} 
where $\phi_{n\vec{k}_0}(\vec{r})=\e^{\i\vec{k}\cdot\vec{r}}u_{n\vec{k}_0}(\vec{r})$ is a Bloch function at the band extremum. 

In the isotropic approximation, the effective mass tensor is replaced by a single averaged effective mass $m^{*}$, resulting in an effective isotropic equation for the envelope function, which is then independent of the index $n$:
\begin{equation}\label{eq:effmasseqn}
[-\frac{\hbar^2}{2m^{*}}\nabla^2-\frac{e^2}{4\pi\epsilon_0\epsilon_r r}-\epsilon]F(\vec{r})=0,
\end{equation}
where $\epsilon_r$ is the relative permittivity of the host. In this paper we will work with this isotropic equation as our starting point. For silicon, $m^*=0.33 m_e$ and $\epsilon_r=11.7$; this leads to a set of scaled atomic units for the hydrogenic impurity problem (length $a_0^*=1.94\ \mathrm{nm}$, energy $\mathrm{Ha}^*=62\ \mathrm{meV}$). For multi-donor systems, the screened Coulomb interaction between electrons $\frac{e^2}{4\pi\epsilon_0\epsilon_r\abs{\vec{r}_1-\vec{r}_2}}$ can be scaled as well. Thus, we have a Hamiltonian in units of $a_0^*$ and the effective Hartree ($\mathrm{Ha}^*$) that reads
\begin{equation}\label{eq:mutlidonors}
\hat{H}=\sum_{i, A}[-\frac{1}{2}\nabla_i^2-\frac{1}{\abs{\vec{r}_i-\vec{R}_A}}]+\sum_{i < j}\frac{1}{\abs{\vec{r}_i-\vec{r}_j}},
\end{equation}
where $A$ runs over all the donor sites; $i$ and $j$ label electrons. To solve this equation, standard molecular \textit{ab initio} computational methods, including CI and TDDFT, can be used to compute excited states.

\subsection{CI and TDDFT methods}\label{subsec: citddft}
%CI
For the CI calculations, we used a specially constructed basis set designed to reproduce within an accuracy of $10^{-5}$ Ha$^*$ the excitation energies of $2s$, $2p_x$, $2p_y$ and $2p_z$ states of a hydrogen atom. We found that 9 Gaussians with exponents ranging from $400/a_0^{*2}$ to $0.008 /a_0^{*2}$ are required, for both $s$ and $p$ symmetries, giving a total basis set of $36$ Gaussians per atom. This basis set was then employed consistently throughout all the CI calculations. CI calculations were performed for a donor pair (DA2) and a uniform three-donor array (DA3) using the Gaussian 09 \cite{gaussian09, nakatsuji1977} and Molpro \cite{molpro1,molpro2, knowles1984, knowles1989} codes.  We used the symmetry-adapted cluster/configuration interaction (SAC-CI) method \cite{nakatsuji1977}, as implemented in Gaussian 09, to calculate the energies and oscillator strengths. We also performed FCI calculations in Molpro \cite{knowles1984, knowles1989}, which gives only the excitation energies: the computation of oscillator strengths within FCI has not been implemented in Molpro, so they were instead estimated using the Gaussian 09 code. The SAC-CI methods implemented in Gaussian 09 can be used to compute the total-spin eigenstates, whereas Molpro produces eigenstates of a given spin projection. FCI calculations can produce accurate excitation energies and yield FCI wave-functions. However, the FCI procedure is limited by the size (the number of electrons and the number of basis functions) of the system as the number of configurations taken into account increases factorially. We have performed the CI calculations for DA2 and DA3 with increments of $\approx 0.07 a_0^*$ ($\approx0.14$ nm, approximately one quarter of silicon lattice constant). 

%TDDFT
As an alternative to CI, TDDFT has been widely used to compute approximately the excited states of molecules and solids \cite{petersilka1996,onida2002} (detailed reviews can be found in Ref.\cite{gross1996}). In TDDFT, the computed excitation energies correspond to the poles of the linear response of the charge density to an external time-dependent stimulus. The linear response of charge densities of the real system is calculated by using the response in a non-interacting reference system, via a formalism similar to the Dyson equation. For our TDDFT calculations, we use an adiabatic hybrid-exchange functional \cite{blyp1, blyp2, blyp3} with the proportion of exact Fock exchange tuned to match the analytical excitation energies of $2s$, $2p_x$, $2p_y$ and $2p_z$ states of a hydrogen atom as accurately as possible; this can be thought of as seeking the best (approximate) cancellation of the self-interaction error in the isolated atom. The optimal proportion of exact Fock exchange was found to be $40 \%$ in order to match the $1s \rightarrow 2sp$ excitation energies of a hydrogen atom (up to $10^{-2}$ Ha$^*$), in comparison with $20 \%$ in the conventional hybrid-exchange functional B3LYP \cite{b3lyp}. We have also tuned the basis set by including more basis functions with smaller gaussian exponents to reproduce the $1s \rightarrow 2sp$ excitation energies, and found that the range for the gaussian exponents is between $400/a_0^{*2}$ to $0.005 /a_0^{*2}$, which is slightly more diffuse than those used in CI calculations though the total number of basis function remains 36 per atom. The basis set and corresponding fine-tuned exchange-correlation functional, implemented in Gaussian 09 code, were then used in all the TDDFT calculations for the excited states of DA2, all the way up to DA10 (uniform ten-donor array). 

We found that in order to represent properly the states at large donor separations, it is important to allow the static DFT solution to find ground states with broken-symmetry form \cite{noodleman}, in which the Kohn-Sham states of opposite spin components are free to localise on different donors.  Although such a solution breaks both the spatial and spin symmetries of the complex, it allows the best approximate representation of the anti-ferromagnetic correlations in the ground state within Kohn-Sham theory \cite{noodleman}.  In practice we find such broken-symmetry configurations are favoured when the atomic separation is greater than approximately $5 \ a_0^*\approx 10\,\mathrm{nm}$ (this can be compared with the experimentally observed Mott transition in three-dimensional doped silicon, where the electrons localize below a density of $3.7\times 10^{18}\,\mathrm{cm}^{-3}$, corresponding to a mean separation of approximately 6.5\,nm) \cite{thomas1981}.  As a result, our TDDFT calculations conserve the total spin projection $M_S$ on the quantization axis, but not the total spin quantum number $S$.

The arrays formed by the uniformly spaced donors are arranged along the $z$-$direction$ throughout the paper, and we discuss all symmetries within the $D_{\infty h }$ point group. We use distance units of nm and energy units of meV throughout. For all the plots of the oscillator strengths, the excitation energies are computed as the energy differences between excited states and the ground state with the same spin, while for the plots of excitation energies, all states are referred to the overall lowest-spin ground state. 

\section{Results}\label{sec: results}
\subsection{Two and three donors: configuration interaction calculations}
\subsubsection{Two donors}
\begin{figure*}[htbp]
%\begin{tabular}{cc}
%\includegraphics[width=9cm,height=5cm]{fig_1a.eps}\\
\includegraphics[width=14cm, height=19cm, trim={0cm 0.2cm 0.8cm 0.2cm},clip]{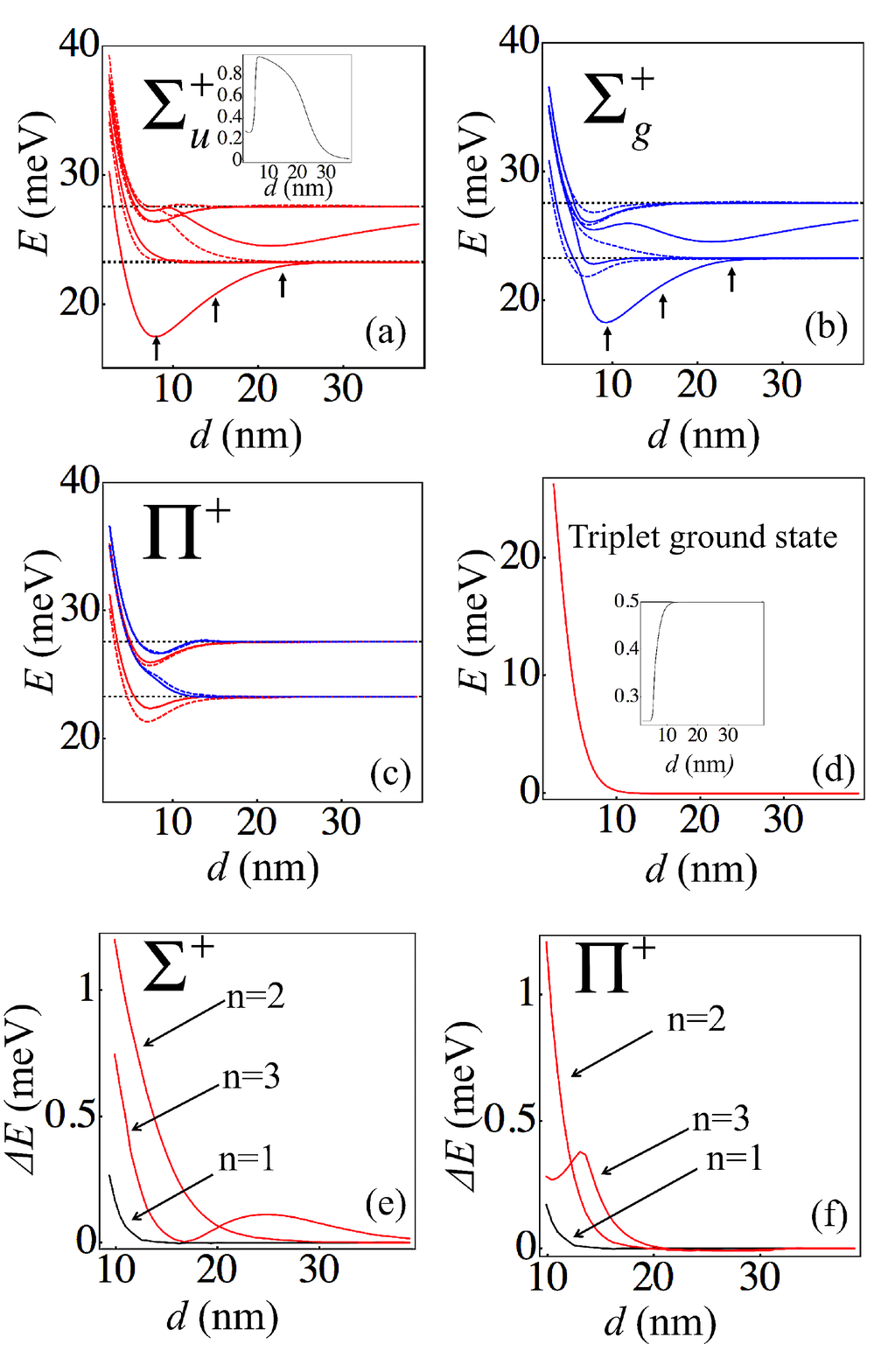}\\
%\textbf{(a)}\\
%\includegraphics[width=8cm,height=7cm]{}\\
%\includegraphics[width=18cm, height=12.35cm, trim={0.cm 0cm 0cm 0cm},clip]{fig_1b.pdf}\\
%\textbf{(b)}\\
%\end{tabular}
\caption{(Colour online.) The excitation energies of DA2 as a function of donor separation, calculated using FCI methods. The first three panels show excitation energies of different spatial symmetries for states converging to $n=2$ and $n=3$ transitions, relative to the overall singlet ground state ($^1\Sigma_{g}^+$): (a) $^1\Sigma_u^+$ and $^3\Sigma_u^+$, (b) $^1\Sigma_g^+$ and $^3\Sigma_g^+$ and (c) $^1\Pi_{u,g}^+$ and $^3\Pi_{u,g}^+$.  Singlet states are indicated by solid lines, triplets by dashed lines.  Odd-parity ($u$) excitations are shown in red, even-parity ($g$) excitations in blue (online)---hence the red (blue) states are accessible by electric-dipole-allowed transitions from the even-parity singlet (odd-parity triplet) ground states.  The inset to (a) shows the CI coefficient of the ionic state in the total wave function of the lowest $^1\Sigma_u^+$ excitation as a function of donor separation. The splitting between the $^1\Sigma_g^+$ singlet and $^3\Sigma_u^+$ triplet ground states within the lowest (1s) manifold as a function of donor separation is shown in (d); the inset is the probability of the double excitation to the $1s$ anti-bonding state in the CI total ground-state wave function, as a function of donor separation. The exchange splittings between corresponding optically accessible excited spin states are shown in (e) and (f): $^1\Sigma_u^+$ and $^3\Sigma_g^+$ excited states (corresponding to excitation from the lowest manifold with $z$-polarization, in red for both $n=2$ and $n=3$ states) are shown in (e), $^1\Pi_u^+$ and $^3\Pi_g^+$ states ($x,y$-polarization, in red)  in (f), along with the ground-state exchange splittings (in black).}\label{fig:1}
\end{figure*}

The ground state of a donor pair has a symmetry of $^1\Sigma_g^+$. In Fig.~\ref{fig:1} we plot the energies of low-lying excited states, with different spatial and spin symmetries ($^{1,3}\Sigma^+_{u,g}$ and $^{1,3}\Pi^+_{u,g}$), as a function of donor separation (singlet in solid curves and triplet in dashed). These excited states converge to the excitations of the $s$ and $p$-states either in the $n=2$ or $n=3$ shells in the limit of isolated donors. The nature of the states is familiar from the previous experimental and theoretical studies of a $\mathrm{H}_2$ molecule \cite{sharp71, kolos}. The four lowest excitation energies of $^1\Sigma_{u}^+$, $^1\Sigma_{g}^+$ and $^1\Pi^+_{u,g}$ symmetries are shown in Fig.\ref{fig:1}(a--c), respectively (in each case two states dissociate to $n=2$ and another two to $n=3$ shells). The singlet excitation energies all rise to $\approx 40$ meV at small separations, where the two donors are strongly coupled, forming a molecular complex. The dominant optical transitions are then between these delocalised molecular orbitals.  

At various points the lowest $^1\Sigma_u^+$ and $^1\Sigma_g^+$ states are formed by different combinations of $1s$, $2s$, and $2p_z$ atomic orbitals \cite{thomas1981, kolos}:  the different regimes are illustrated by the arrows in Fig.\ref{fig:1}(a--b). At small separations (the leftmost arrow marking the minimum excitation energy of $\approx 17$ meV, $d<7\,\mathrm{nm}$) the excitations are predominantly between molecular orbitals, converging to the single-electron $2p\sigma$  ($^1\Sigma_u^+$) and $2s\sigma$ ($^1\Sigma_g^+$) excitations in the united-atom limit \cite{sharp71}.  Near the centre arrows ($7\,\mathrm{nm}\le d\le 22\,\mathrm{nm}$) these lowest singlet excitations correspond closely to the ionic (or charge-transfer) excited state, which can also be thought of as arising from the transition between the $1s\sigma$ (bonding) and $1s\sigma^*$ (anti-bonding) states. The excitation energy has a minimum of $\approx 17$ meV at an inter-donor distance of $\approx 7 $ nm. At $d\approx 22$ nm (centre arrow), there is a transition where the ionic state anti-crosses with the $1s\rightarrow 2sp$ transition; at larger separations the lowest excitation has a predominantly single-atom $1s\rightarrow 2sp$ character, while the charge-transfer transition increases further in energy towards the $1s\rightarrow3sp$ excited states (rightmost arrow).  Meanwhile four further $1s\rightarrow 2sp$ transitions ($^{1}\Sigma^+_{u, g}$ and $^{1}\Pi^+_{u, g}$) persist with their energies almost unaffected as long as the donor separation is larger than $\approx12\,\mathrm{nm}$, below which the hybridization between orbitals in the $n=2$ shell on different atoms starts to become significant. The $^1\Sigma_g^+$ state becomes the $3s\sigma$ state of the united atom, the $^1\Sigma_u^+$ state leads to the united-atom $3p\sigma$ excitation, while the $^1\Pi_{u}^+$ drops briefly in energy to form the $2p\pi$ excitation and the $^1\Pi_{g}^+$ rises steeply to form the $3d\pi$ excitation. The upper band of $^1\Sigma_{u,g}^+$ charge-transfer states (beyond the anti-crossing with the single-atom $n$ = 2 transition) transforms below 22\,nm (rightmost arrow) into a combination of $2s$, $2p_z$, $3s$ and $3p_z$ atomic excitations, which rises gradually in energy before splitting at approximately $13\,\mathrm{nm}$ when reducing donor separations, partially containing an ionic-state nature. The upper ($\Sigma_u^+$) branch then crosses the other $1s\rightarrow3sp$ transitions at a donor separation of approximately $11\,\mathrm{nm}$. The $^1\Pi_g^+$ ($^1\Pi_u^+$) states correspond to the $1s\rightarrow 2p_{xy}$  and $1s\rightarrow 3p_{xy}$ transitions. 

%In Fig.\ref{fig:1}(d), we show the excitation energies of the triplet ground state ($^3\Sigma_u^+$) relative to the singlet ground state ($^1\Sigma_g^+$) as a function of donor distance, which is the exchange splitting within the ground-state manifold. 
The corresponding low-lying triplet excited states with $^3\Sigma_{u}^+$, $^3\Sigma_{g}^+$ and $^3\Pi^+_{u,g}$ symmetries are shown (relative to the singlet ground state $^1\Sigma_g^+$) as the dashed curves in Fig.\ref{fig:1}(a--c), respectively.  
There is no charge-transfer state (because of the Exclusion Principle) and the six transitions converging to $n=2$ for separated atoms remain almost degenerate down to $d\approx 15\,\mathrm{nm}$. At this separation, when reducing inter-donor distance, one of the $^3\Sigma_u^+$ states rises sharply before becoming the excitation to the $4f\sigma$ orbital in the united atom, while one of the $^3\Sigma_g^+$ states drops sharply to form the $2s\sigma$ excitation in the united atom. 

Further insight into the nature of the states can be obtained from their compositions in terms of molecular-orbital excitations. The inset to Fig.\ref{fig:1}(a) shows the CI coefficient of the ionic state, arising from the $\sigma_g(1s)\rightarrow\sigma_u^*(1s)$ (bonding to anti-bonding) transition, in the total wave function of the first $^1\Sigma_u^+$ excited state; the coefficient peaks at a donor distance of $\approx 7$ nm, near the minimum excitation energy.  At smaller separations the lowest excitation has only partially ionic character, while at larger separations it remains substantially ionic until the anti-crossing at approximately $d=22\,\mathrm{nm}$. Similarly, the inset of Fig.\ref{fig:1}(d) shows the probability (the square of the CI coefficient)  of the doubly excited configuration to the $1s$ anti-bonding state in the ground-state total wave function; this is a measure of the correlation effects in the ground state that correct the single-particle picture. This probability increases sharply from $\frac{1}{4}$ to $\frac{1}{2}$ at a donor separation of $\approx 5$ nm, corresponding to the evolution from a delocalised molecular-orbital excitation to a localised state with one electron per donor. We can take the separation where the probability reaches $\frac{1}{2}$ ($\approx 10$ nm) as an indicator of the location of the Mott transition; this is in reasonable agreement with the experimentally observed transition density \cite{thomas1981}, as well as with the onset of the broken-symmetry ground state in DFT that will be shown later. 

A quantity of particular interest for the optical control of spin couplings, and ultimately for the development of optically controlled quantum gates \cite{sfg, filidou2012}, is the triplet-singlet exchange splitting as a function of donor separations. The exchange splitting here is defined as the energy difference between corresponding triplet and singlet states (positive sign means antiferromagnetic, negative sign ferromagnetic).  Care must be taken to compare pairs of states that are orbitally similar, and for separations below about 10\,nm, intersections among the excited states make it difficult or impossible to define an exchange interaction properly.  Fig.~\ref{fig:1}(d) shows the exchange splitting between the $^1\Sigma_g^+$ singlet ground state and the $^3\Sigma_u^+$ triplet ground state.  For the exchange splitting in the excited state, supposing we start from a general spin state in the manifold of states dissociating to ground-state atoms; this will be a linear combination of $^1\Sigma_g^+$ and $^3\Sigma_u^+$ ground states. With light polarised along the donor pair axis ($z$-direction) we will excite to a corresponding combination of $^1\Sigma_u^+$ and $^3\Sigma_g^+$, while with light polarised perpendicular to the axis we will make a combination of $^1\Pi_u^+$ and $^3\Pi_g^+$. The exchange splittings between the appropriate $^1\Sigma_u^+$ and $^3\Sigma_g^+$ ($^1\Pi_u^+$ and $^3\Pi_g^+$) states for a single electron excited to $n=2$ and $n=3$ shells, are shown in Fig.\ref{fig:1}(e) (Fig.\ref{fig:1}(f)). Notice that the splittings are generally anti-ferromagnetic in sign and, as expected, considerably larger and longer range than the exchange splitting in the ground-state manifold (shown again for comparison). This coupling could be used to realize two-qubit quantum gate operations by using optically excited donor states, as mentioned previously \cite{sfg}.  

%As expected, the former is larger but decay faster as compared to the latter one. It is noteworthy that the exchange coupling in the $\Pi_u^+$ symmetry.

%discuss donor density for MIT

%Meanwhile pairs of doubly degenerate $\Pi_u$ and $\Pi_g$ states, formed by the $2p_{x,y}$ and $3p_{x,y}$ orbitals perpendicular to the donor array axis ($z$-axis), also converge to the 2sp and 3sp atomic energies respectively, and do not interact with the charge-transfer band.  Similarly the excitation energies in the triplet sector from the ground state (Fig.\ref{fig:1}(b)  show no ionic state (forbidden by spin symmetry), but only atomic excitation at large separations, with splitting between the different symmetry components setting in around $D=15\,\mathrm{nm}$.

\begin{figure*}[htbp]
%\begin{tabular}{c}
%\includegraphics[width=8cm,height=7cm]{}\\
%\includegraphics[width=8.5cm, height=5.5cm, trim={0.0cm 0cm 0.cm 0.0cm},clip]{fig_2a.pdf}\\
%\textbf{(a)}\\
%\includegraphics[width=8cm,height=7cm]{}\\
\includegraphics[width=15cm, height=22cm, trim={0.0cm 0cm 1cm 0.5cm},clip]{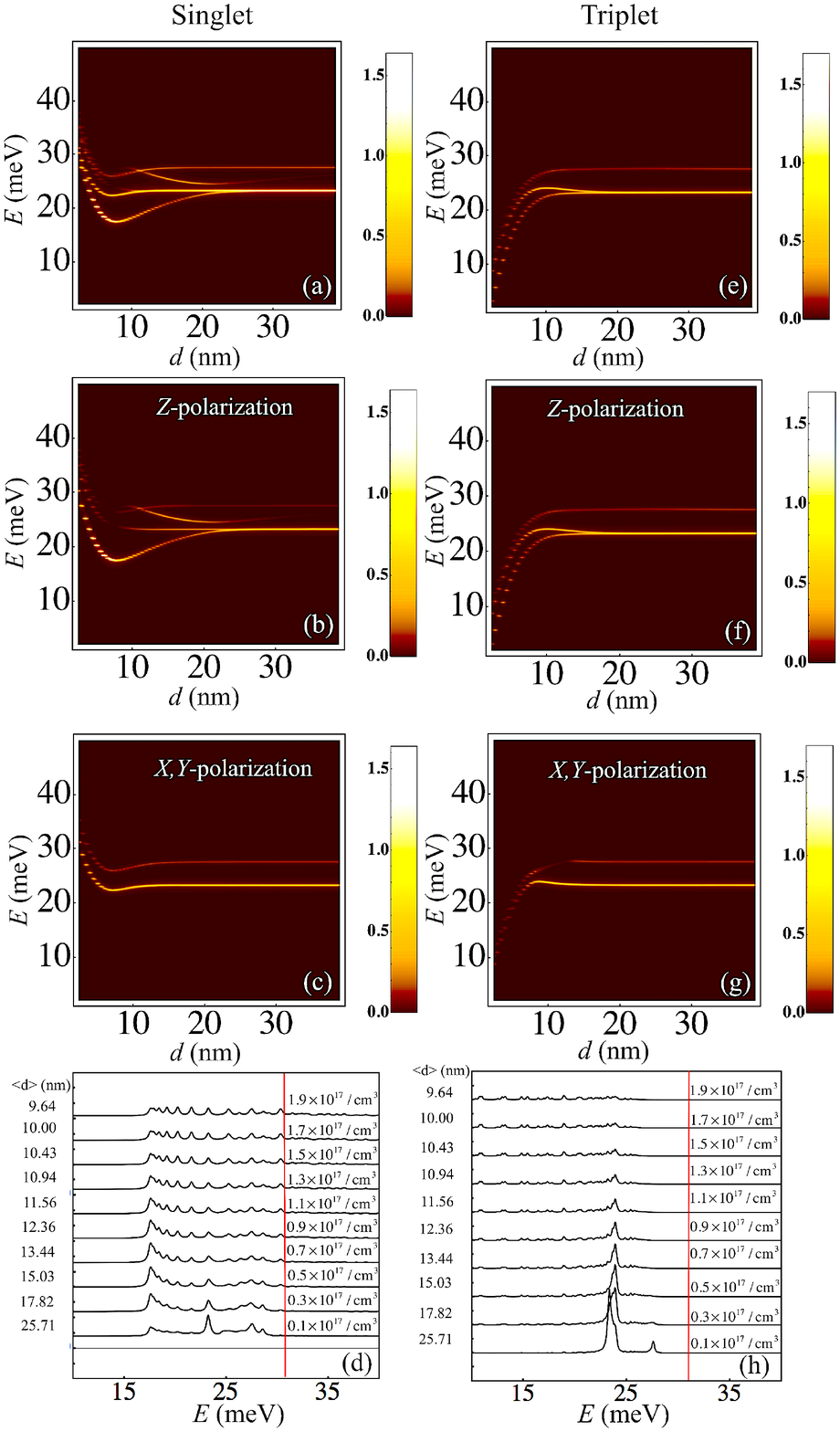}
%\textbf{(b)}\\
%\end{tabular}
\caption{(Colour online.) The optical absorption of DA2 as a function of excitation energy and donor separation. The broadened oscillator strengths for all singlet excitations and all polarizations are shown in (a). The oscillator strengths for polarized excitation along $z$-, and $x,y$-directions are shown in (b-c). The corresponding plots for the triplet excitations are shown in (e-g). The statistically averaged values of these oscillator strengths for the singlet and triplet sectors, according to the random distribution of the first-nearest-neighbours, are shown in (d) and (h) respectively as a function of excitation energy, for a set of donor densities ($0.1\times10^{17}$cm$^{-3}$ to $1.9\times10^{17}$cm$^{-3}$ with $0.2\times10^{17}$cm$^{-3}$ increments). We also show the corresponding mean donor distance for each density. The red vertical lines in (d) and (h) correspond to the donor ground-state ionization energy ($\frac{1}{2}\mathrm{Ha}^*\approx 31$ meV) within the EMT approximation.  Only those states converging to $n=2$ or $n=3$ transitions at large separations are included.}\label{fig:2}
\end{figure*}

%\subsection{Average excitation spectra}
Figure \ref{fig:2}(a) shows the total oscillator strengths of transitions from the overall (singlet) ground state as a function of the inter-donor distance and the excitation energy, obtained by using a Lorenz-type broadening with a half-width $0.1$ meV (centred at the excitation energies, the height being the oscillator strength) \cite{chick2017}. This shows clearly the positions of optically accessible states for different separations: at large separations the spectrum is dominated by the $1s\rightarrow 2p$ and $1s\rightarrow 3p$ atomic transitions, while at separations below approximately 20\,nm the charge-transfer band and the corresponding anticrossed (mainly $3sp$) state also contribute strongly to the oscillator strength. The ionic excitation is at $\approx 17 $ meV, which is comparable to the experimental results in Ref.\cite{thomas1981} if the correction for the lowering of the $1s(A)$ state from the central cell and inter-valley effects is included ($\approx 14$ meV). The charge-transfer state makes a negligible contribution to the oscillator strength beyond the anti-crossing with the $2sp$ excitations ($d >$22\,nm). In Fig.\ref{fig:2}(b-c), we show the contributions to the oscillator strength along $z$- and $x$- (or $y$-) directions for all the optically accessible states, respectively, which correspond to $\Sigma$- and $\Pi$-transitions. Fig.\ref{fig:2}(d) shows the averaged oscillator strength as a function of energy for a distribution of nearest-neighbour distances corresponding to random donor placements with a range of donor densities, where donor pairs are important \cite{thomas1981, wu2008, wu2007}. This gives an approximate description of the absorption of a randomly doped crystal under the assumption that pairwise interactions dominate (at least, in the low-energy regime where transitions to other bound states dominate; our basis set is not designed to describe bound-to-continuum transitions at higher energies). For diluted systems the main peak is at the $1s\rightarrow2sp$ transition energy ($\approx 23$ meV), while as the donor density increases (from $0.3\times10^{17}\mathrm{cm}^{-3}$ to $1.7\times10^{17}\,\mathrm{cm}^{-3}$) the lower-energy ionic-state excitations strengthen. For each density, we also show the mean donor distance in the continuum limit  ($\langle d\rangle = \Gamma(\frac{4}{3})(\frac{4\pi n}{3})^{-\frac{1}{3}}$)  on the left-hand side of Fig.\ref{fig:2}(d) and (h). In Fig.\ref{fig:2}(e-h), we show the corresponding plots for triplet excited states as for the singlet sector. Notice that for triplet states the averaged oscillator strengths suggest that the atomic transition is dominant; by contrast, the oscillator strengths of low-energy triplet states at small separations are weak.  

\subsubsection{Three donors}
\begin{figure*}[htbp]
%\begin{tabular}{c}
%\includegraphics[width=8cm,height=7cm]{}\\
%\includegraphics[width=8.5cm, height=6cm, trim={0.cm 0.0cm 0.1cm 0.cm},clip]{fig_3a.pdf}\\
%\textbf{(a)}\\
%\includegraphics[width=8cm,height=7cm]{}\\
%\includegraphics[width=8.5cm, height=6cm, trim={0.cm 0.0cm 0.1cm 0.2cm},clip]{fig_3b.pdf}\\
%\textbf{(b)}\\
\includegraphics[width=18cm, height=14cm, trim={4.0cm 0.0cm 0.0cm 0cm},clip]{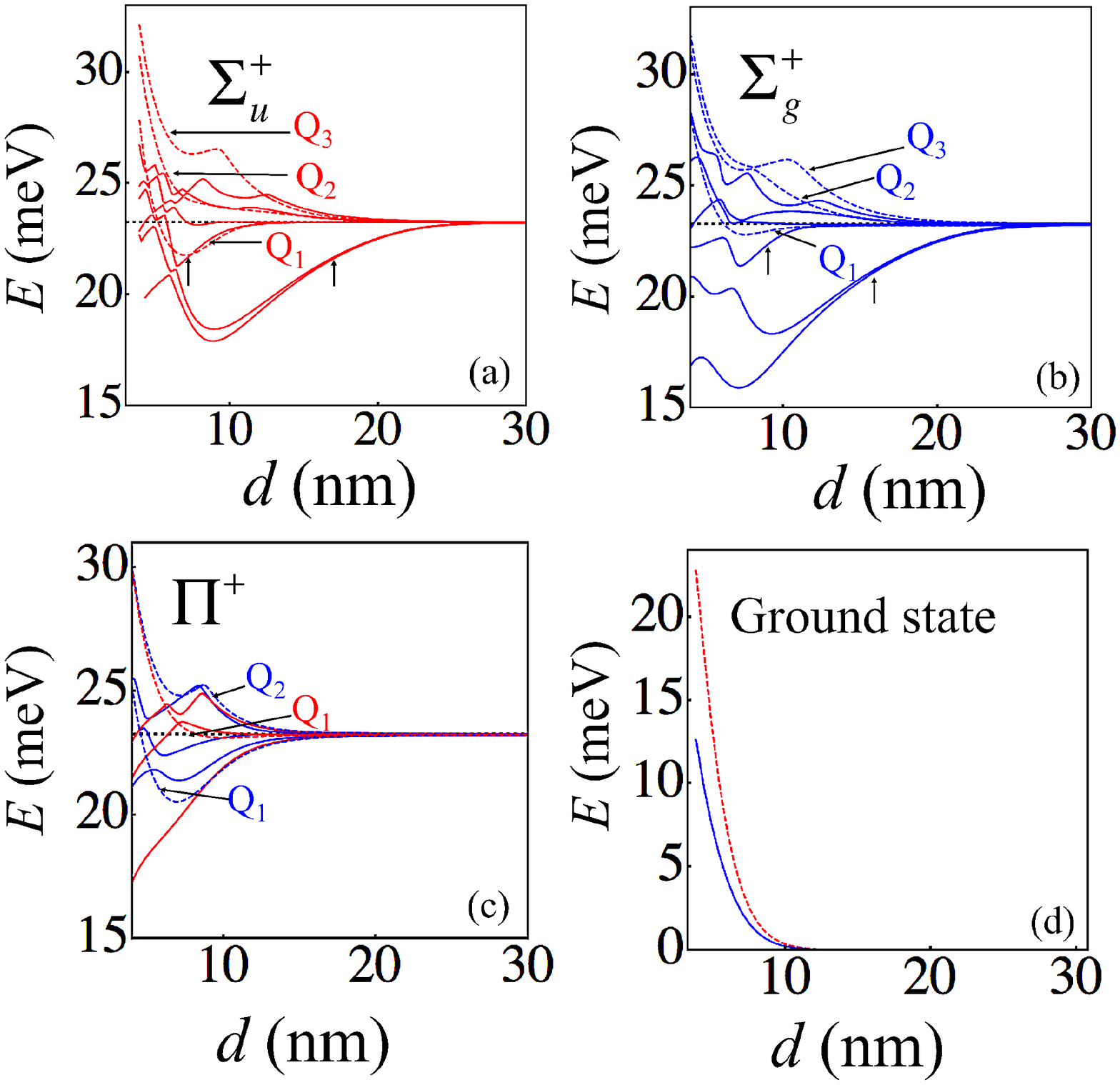}\\
%\textbf{(c)}\\
%\end{tabular}
\caption{(Colour online.) Excitation energies of DA3 above the ground state as a function of donor separation, computed using FCI methods, are shown.   States converging to an excited atom with $n=2$ are shown for different spatial symmetries:  (a) $\Sigma_u^+$, (b)  $\Sigma_g^+$, and (c) $\Pi_{u,g}^+$.  Full lines are doublet (low-spin) states, dashed lines (labeled in order to clarify exchange splittings) are quartet (high-spin) states; odd-parity ($u$) states are shown in red, even-parity ($g$) in blue.  For the $\Sigma$ symmetries, the vertical arrows point to the ionic states with first nearest-neighbour separation (right vertical arrows) and second nearest-neighbour separation (left vertical arrows).  (d) shows the excitation energies of $^4\Sigma_u^+$ and $^2\Sigma_g^+$ states in the lowest manifold of states that dissociate to isolate atoms in the ground state. All the quartet states in (a-c) have been labeled for further energy comparison (adopting the same colour scheme as others).}\label{fig:3}
\end{figure*}

Figure \ref{fig:3}(a--d) show the excitation energies of a line of three uniformly distributed donors relative to the ground state ($^2\Sigma_u^+$ symmetry) for low-lying states with $S=\frac{1}{2}$ (doublet) and $S=\frac32$ (quartet), as a function of inter-donor distance. Low-spin (high-spin) states are plotted by solid (dashed) curves. In total there are 6 $^2\Sigma_{u}^+$, 6 $^2\Sigma_g^+$, 3 $^2\Pi_g^+$, and 3 $^2\Pi_u^+$ doublet states, 3 $^4\Sigma_{u}^+$, 3 $^4\Sigma_g^+$, 2 $^4\Pi_g^+$, and 1 $^4\Pi_u^+$ quartet states, converging to the isolated donor $n=2$ transitions in the limit of large separations as shown in Fig.\ref{fig:3}(a-c), and the same numbers converging to $n=3$ (not shown, for clarity). Note that since the ground state has odd parity the optically allowed transitions are now to even-parity states (blue curves). In addition, there is a manifold of two low-lying excited states within the $1s$ subspace, one $^2\Sigma_g^+$ and one $^4\Sigma_u^+$, which is the quartet ground state, as shown in Fig.\ref{fig:3}(d). At large separations these two low-lying states converge to spin excited states of the three-spin Heisenberg chain.  At smaller separations they become single-particle excitations into the two excited molecular orbitals formed by linear combinations of the donor $1s$ orbitals. 

We now find two different types of ionic states, indicated by the vertical arrows in Fig.~\ref{fig:3}(a) and (b): one branch (right-hand vertical arrow) splits off from the $n=2$ transitions at approximately $d=22\,\mathrm{nm}$ as in the two-donor case, and corresponds to an ionic state on first nearest neighbours. The other is at a higher energy (left-hand vertical arrow) anti-crossing the $n=2$ atomic transition at $d \approx 12$ nm, splitting from the $n=3$ transitions (not shown) at approximately $d=25\,\mathrm{nm}$, and has the electron and hole located on second nearest neighbours. These two types of ionic states anti-cross each other at $d \approx 5$ nm, where the anti-crossing gap of the $^2\Sigma_g^+$ symmetry is much larger than that of the $^2\Sigma_u^+$. 

Fig.\ref{fig:3}(c) shows the low-lying states of $\Pi$ symmetry (all doubly orbitally degenerate, in the absence of spin-orbit coupling); as for two donors, they all converge to the $n=2$ energy at large separations, and significant interactions between them on this scale are visible only below separations approximately $d\approx13\,\mathrm{nm}$. 
%XXX Include plots of doublet-quartet splittings XXX XXX Include oscillator strengths XXX
\begin{figure*}[htbp]
%\%begin{tabular}{c}
%\includegraphics[width=8cm,height=7cm]{}\\
\includegraphics[width=15cm, height=15cm, trim={0.1cm 0cm 0.5cm 0.5cm},clip]{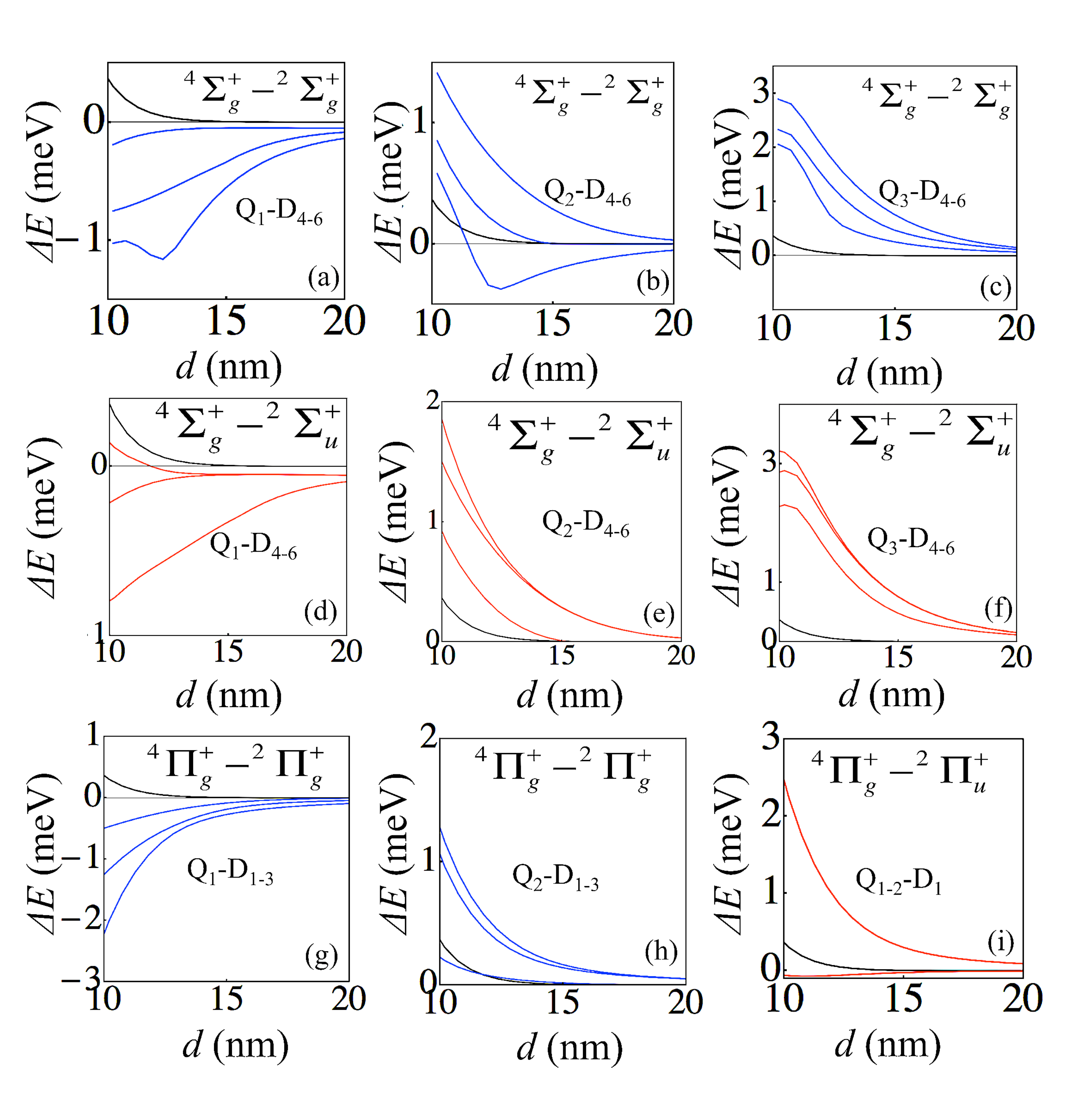}
%\textbf{(a)}\\
%\includegraphics[width=8cm,height=7cm]{}\\
%\includegraphics[width=8.5cm, height=5.5cm, trim={0.cm 0cm 0.cm 0cm},clip]{fig_4b.pdf}\\
%\textbf{(b)}\\
%\end{tabular}
\caption{(Colour online.) The exchange splittings between quartet and doublet states in the excited-state manifolds for DA3 as a function of donor spacing:  (a--c) show splittings between states Q$_1$--Q$_3$ respectively of $^4\Sigma_g^+$ symmetry (see Fig.~\ref{fig:3}) and states D$_4$--D$_6$ (excluding the ionic states D$_1$ to D$_3$) of $^2\Sigma_g^+$ symmetry (all are excited states converging to $n=2$ excitations), while (d--f) show splittings between corresponding states of $^4\Sigma_g^+$ and $^2\Sigma_u^+$ symmetries.  These excitations can all be accessed from the ground-state manifold with polarisation along the $z-$direction. Energy differences between even-parity states are in blue, even and odd states in red. (g--h) show similar splittings between $^4\Pi_g^+$ and $^2\Pi_g^+$ states (produced by excitation along $x-$ and $y-$directions), while (i) shows splittings between $^4\Pi_g^+$  and $^2\Pi_u^+$ states; the colour scheme is as for polarisation along $z$-direction. The lowest quartet-doublet splitting in the ground-state manifold is also plotted  in (a) -(i) for comparison (black curves). Quartet states are ordered from low to high energy as labeled in Fig.\ref{fig:3} (similarly for the doublet states). }\label{fig:4}
\end{figure*}
\begin{figure*}[htbp]
\includegraphics[width=15cm, height=6cm, trim={0cm 0cm 0cm 0cm},clip]{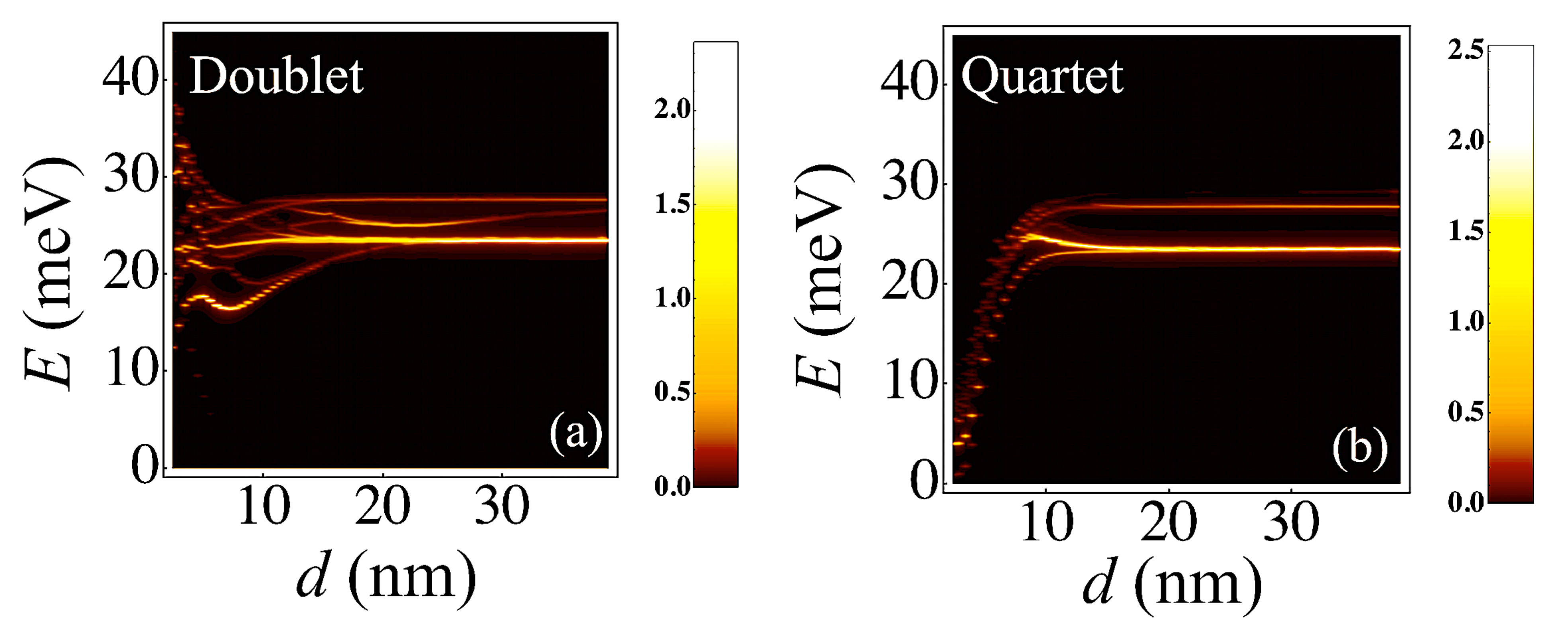}
%\textbf{(e)}&\\
%
%\end{tabular}
\caption{(Colour online.) The broadened oscillator strengths for the doublet and quartet excited states of DA3 are shown in (a) and (b), respectively. For doublet, the ionic states are dominant at the mid-range for donor separation, whereas the atomic transitions for the long-range. For quartet states, the atomic transitions are clearly important.}\label{fig:5}
\end{figure*}
Figure \ref{fig:4} (a--i) show the relevant exchange splittings for excited states that can be accessed by optically allowed transitions from linear combinations of states in the $1s$ low-energy subspace (the $^2\Sigma_u^+$ ground state and the low-lying $^2\Sigma_g^+$ and $^4\Sigma_u^+$ excitations). The relevant excited states therefore include $^2\Sigma_g^+$, $^2\Sigma_u^+$ and $^4\Sigma_g^+$ (for polarization along the $z$-direction), and $^2\Pi_g^+$, $^2\Pi_u^+$ and $^4\Pi_g^+$ (polarization along $x$- or $y$-direction). Fig. \ref{fig:4}a--c (d--f) show the splittings between $^2\Sigma_g^+$ ($^2\Sigma_u^+$) states and Q$_1$-Q$_3$ of the $^4\Sigma_g^+$ states (see also Fig.~\ref{fig:3}), respectively. Fig.\ref{fig:4} g--h(i) show the corresponding splittings for excitation polarised along $x$- or $y$-direction, between $^2\Pi_g^+$ ($^2\Pi_u^+$) and $^4\Pi_g^+$ states. As in the donor pair (Fig.~\ref{fig:1}), the exchange splittings are much larger in the excited states than in the-ground state manifold, indicating the potential of optical excitations to control the exchange interaction, and hence implement spin-based quantum gate operations.

% the calculations for the maximum of quartet-doublet energy splittings in all the symmetries are shown. They are the strongest coupling once the system is excited either from a doublet or quartet ground state. These values also correspond to the shortest times ($t = \frac{\hbar}{E}$) needed for an optically-controlled quantum gate operation \cite{wu2007}. To insure a negligible ground-state exchange, the inter-donor distance should be at least $20$ nm, where the maximum of the excited-state exchange interaction, when one donor is excited, is $\approx 3$ meV, corresponding to $\approx 1.5$ ps. This time scale can insure a quantum operation before the donor excited state relaxes (the relaxation time for an excited state in donors is typically in the order of $100$ ps). 
 
In Fig.~\ref{fig:5}(a--b), we show the broadened oscillator strengths in the three-donor system (DA3) as a function of donor separation and excitation energy, from the doublet ground state $^2\Sigma_u^+$ in (a) and the quartet ground state $^4\Sigma_u^+$ in (b).  At a donor separation of $\approx 7$ nm, the ionic states dominate the doublet optical absorption (to a greater extent than for two donors): the lowest charge-transfer transition is the strongest, while the upper one shows signs of anti-crossing with the longer-range charge transfer state at approximately $d=6\,\mathrm{nm}$.  At long range the intra-donor excitation dominates, as in the two-donor case, though the charge-transfer state is more visible than for two donors.  The absorption is similar to that of donor pair, the main differences being the splitting of the $n=2$ excitation at separations around 10\,nm and additional low-energy quartet excitations appearing at small separations.

\subsection{Longer arrays: TDDFT calculations}
\subsubsection{Benchmarking for two and three donors}
\begin{figure*}[htbp]
%\begin{tabular}{c c}
%
%\includegraphics[width=9cm, height=6cm, trim={0.2cm 0.2cm 0.2cm 0.2cm},clip]{fig_4a.pdf}&\includegraphics[width=9cm, height=6cm, trim={0.cm 0cm 0.cm 0cm},clip]{fig_4b.pdf}\\
%\textbf{(a)}&\textbf{(f)}\\
%\includegraphics[width=9cm, height=6cm, trim={0.cm 0cm 0.cm 0cm},clip]{fig_5a.pdf}&\includegraphics[width=9cm, height=6cm, trim={0.cm 0cm 0.cm 0cm},clip]{fig_5b.pdf}\\
%\textbf{(b)}&\textbf{(g)}\\
%
%\includegraphics[width=9cm, height=6cm, trim={0.cm 0cm 0.cm 0cm},clip]{fig_5c.pdf}&\includegraphics[width=9cm, height=6cm, trim={0.cm 0cm 0.cm 0cm},clip]{fig_5d.pdf}\\
%\textbf{(c)}&\textbf{(h)}\\
%
%\includegraphics[width=9cm, height=6cm, trim={0.cm 0cm 0.cm 0cm},clip]{fig_5e.pdf}&\includegraphics[width=9cm, height=6cm, trim={0.cm 0cm 0.cm 0cm},clip]{fig_5f.pdf}\\
%\textbf{(d)}&\textbf{(i)}\\
%
\includegraphics[width=12cm, height=25cm, trim={3cm 0cm 3cm 0cm},clip]{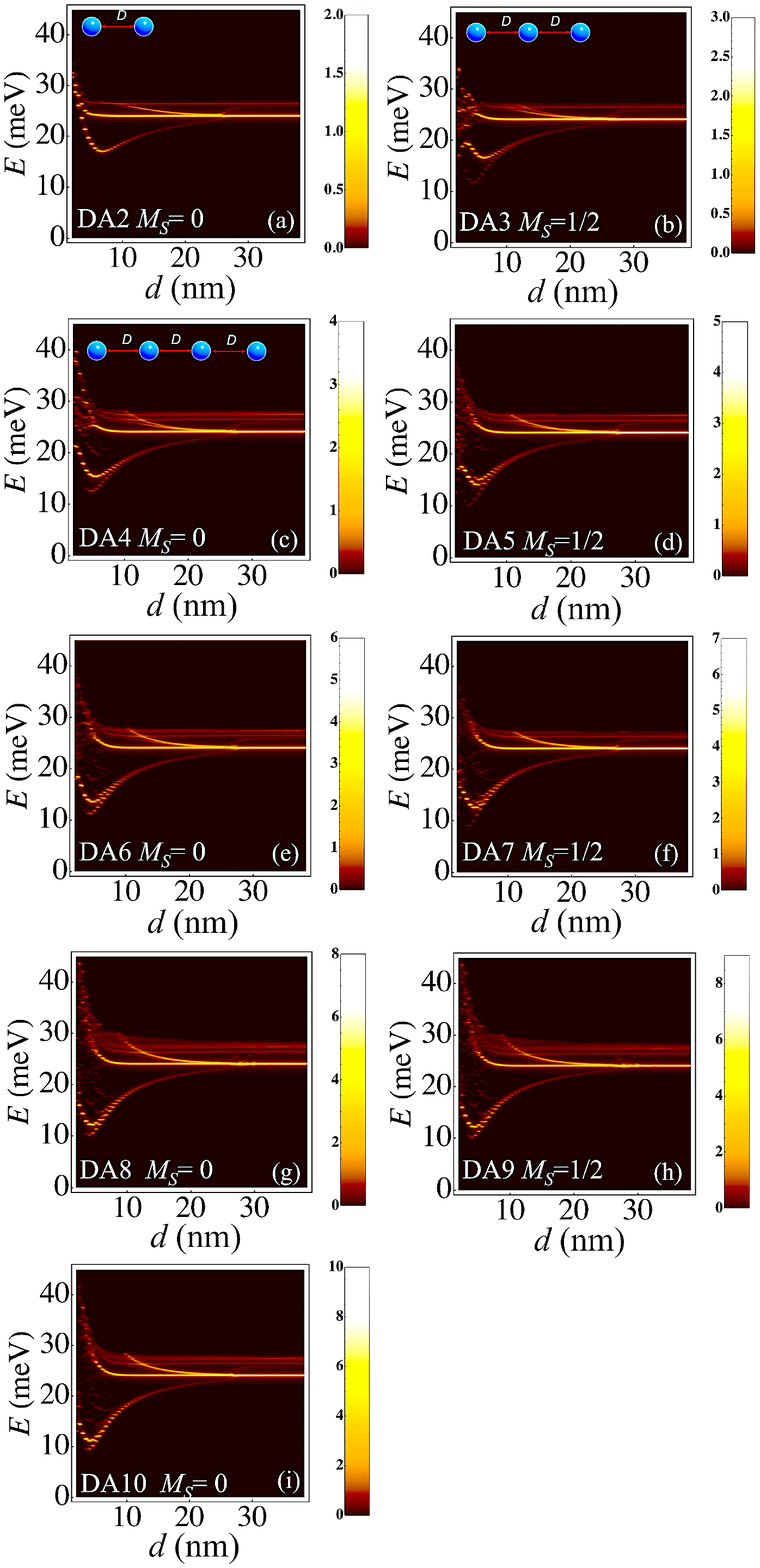}
%\textbf{(e)}&\\
%
%\end{tabular}
\caption{(Colour online.) The broadened oscillator strengths as a function of excitation energy and donor separation for arrays of different sizes from DA2 up to DA10 as calculated in TDDFT are shown in (a) - (i), respectively. Notice that they share generic features: molecular transitions for short separations, charge-transfer bands in the mid-range, and atomic transitions at large separations. The lowest excitation energy falls to $\approx 10$ meV (ionic state, corresponding to a wave length of $\approx 60\ \mu \mathrm{m}$) as the number of donors increases.}\label{fig:6}
\end{figure*}

In addition to CI calculations, we have also performed TDDFT calculations for lines of two to ten donors. 
%\subsubsection{ DA2 and DA3}
For benchmarking, we first show the optically accessible excitation energies as a function of inter-donor distance for DA2 and DA3 in Fig.~\ref{fig:6}(a) and (b). The TDDFT results agree qualitatively with the CI calculations shown for DA2 in Fig.\ref{fig:2}(a) and for DA3 in Fig.\ref{fig:4}(j), correctly capturing the contribution from the ionic states and producing the correct long-distance limits (governed by intra-donor $n=2$ and $n=3$ excitations).  The main qualitative discrepancy is the failure of TDDFT to capture the minimum in the $n=2$ excitation (Fig.\ref{fig:2}(a)) as a function of separation, which could arise from higher-order configurations. In Table~\ref{table:compare-CI-TDDFT}, we compare quantitatively the excitation energies and oscillator strengths for three different separations: one near the minimum in the charge-transfer band at 7\,nm, one in the region dominated by the charge transfer band at 15\,nm, and one in the long-distance regime dominated by intra-donor excitations at 25\,nm.  Quantitatively, TDDFT is found to overestimate the oscillator strengths for the $n=2$ transitions at approximately $d=7\,\mathrm{nm}$, but to underestimate them at larger separations. TDDFT also tends to underestimate the strength of the charge-transfer transition, by up to a factor of two. 

\begin{table*}
\begin{tabular}{ccccccc}
\hline
$d$ (nm)&Transition&Method&\multicolumn{4}{c}{Donor Array} \\
%\cline{4-7}
\hline
&&&\multicolumn{2}{c}{DA2} & \multicolumn{2}{c}{DA3}\\
%\hline
&&&$E$ (meV)&$f$ & $E$ (meV)&$f$ \\
\hline
7&CT&CI&17.59&0.44&16.50&0.54\\
&&TDDFT&17.11&0.21&16.78&0.33\\
&$1s\rightarrow 2sp$&CI&24.08&0.00&21.88&0.01\\
&&TDDFT&23.92&0.04&24.08&0.03\\
\hline
15&CT&CI&20.86&0.08&21.52&0.03\\
&&TDDFT&20.99&0.03&21.08&0.03\\
&$1s\rightarrow 2sp$&CI&23.20&0.10&23.37&0.12\\
&&TDDFT&23.98&0.07&24.00&0.08\\
\hline
25&$1s\rightarrow 2sp$&CI&23.13&0.18&23.38&0.34\\
&&TDDFT&24.34&0.13&24.24&0.19\\
\hline
\end{tabular}
\caption{\label{table: compare-CI-TDDFT}Comparison of selected results for excitation energy $E$ and oscillator strength $f$ of optical transitions from the low-spin ground state, using FCI and TDDFT approaches for DA2 and DA3 at three different separations.   The different transitions are the intra-donor $1s\rightarrow 2sp$ transition and the charge-transfer (CT) transition.}\label{table:compare-CI-TDDFT} 
\end{table*}
%XXX Over what range of energies are we claiming this figure is accurate? XXX
 %\subsubsection{ DA4 - DA10}

\subsubsection{Four to ten donors}
Figure \ref{fig:6}(c--i) show the optically accessible excitations within TDDFT for states of uniform DA4 - DA10 lines, showing the lowest projections of total spin ($M_S=0$ for even chains, $M_S=1/2$ for odd chains). They share a number of qualitative features with one another and with the shorter chains described previously. All have a band of strong absorption at approximately 24\,meV (the $n=2$ intra-donor excitation), an energy that is almost constant down to $d\approx 8\,\mathrm{nm}$, where it starts to rise sharply.  Correspondingly, the flat bands at approximately 28\,meV and 30\,meV are the excitations to $n=3$ and $n=4$ states, split by the incompleteness of the basis and (more importantly) by the incomplete cancellation of self-interaction errors. The band of 'ionic' states corresponding to excitation across the Mott-Hubbard gap also features prominently in all the chains, as it does in the cases of DA2 and DA3 described previously.  For separations $7\,\mathrm{nm}\le d\le 22\,\mathrm{nm}$ the states corresponding to nearest-neighbour excitation are approximately degenerate; below 7\,nm they start to become split by hopping interactions between the donors, similar to those that produce splittings between the $\Sigma_u^+$ and $\Sigma_g^+$ components in the two-donor case (see Fig.~\ref{fig:1}(a)).  At still smaller separations these excitations transform adiabatically into the single-particle excitations from the $1s$ bonding state to the corresponding anti-bonding states; at large separations (above 22\,nm) they merge into the $n=2$ orbital excitation, although the details of the anti-crossing observed in the two- and three-donor CI results are not quantitatively reproduced.  The minimum excitation energy in this ionic band drops to $\approx 10$ meV for long chains.  A sequence of further charge-transfer exciton bands with larger electron-hole separations is expected (as suggested by the left arrow in Fig.~\ref{fig:3}(a) for DA3) but the oscillator strengths for these are exponentially suppressed because of the large charge separations and they are therefore not visible in these plots.

There is also a band of excitations at lower energies, below the Mott-Hubbard gap.  At large separations these correspond to the spin excitations of the Heisenberg spin chain; they have very small charge character and correspondingly negligible electric dipole matrix elements with the ground state, and hence would be invisible on the colour plots; TDDFT does not find them for the lowest spin sectors shown here.  As the separation drops (and the ratio $t/U$ in the corresponding effective Hubbard model rises) these excitations acquire an increasing charge character and split as a result of the increasing inter-donor hopping.  Eventually the highest-lying members of this manifold are expected to merge with the inter-site exciton band at separations where the Mott gap closes. The broken-symmetry ground state come forms above an inter-donor distance of $\approx 10$ nm, which is consistent with that predicted by CI calculations.

\section{Conclusion}\label{sec: conclusion}
We have computed the excitation energies and optical response for one-dimensional donor arrays in silicon with up to 10 dopants within the spherical band approximation. We include a full description of intra- and inter-donor correlation through our CI calculations on small systems (2- and 3-donor arrays), and an approximate description of those correlations through TDDFT calculations for larger systems (from 4- up to 10-donor arrays). We find that these correlations are very important in determining the THz optical spectrum of donors in silicon, giving rise to features such as ionic excited states, which cannot be well described by single-particle molecular energy levels. 

The smallest optically accessible excitation energies within the lowest spin configuration originate from the inter-donor ionic charge-transfer state, which becomes dominant at a donor separation of $\approx 5$ to $10$ nm, corresponding to a donor density $\approx 3 \times 10^{17}$ to $8\times 10^{18} \mathrm{cm}^{-3}$. This donor density region can be well described by a donor-pair model \cite{thomas1981}. The donor-separation range where the ionic states are important extends to $\approx 30$ nm (corresponding to a donor density $\approx 3.7 \times 10^{16}  \mathrm{cm}^{-3}$). At longer range ($> 40$ nm, corresponding to $< 1.6 \times 10^{16} \mathrm{cm}^{-3}$), intra-donor excitations dominate optical transitions. In contrast, at small donor distances (typically smaller than $5$ nm) we have a molecular picture for the excitations. 

The work presented here treats the molecular-type, charge-transfer, and atomic excited states on the same footing. The first two of these features have been seen in a previous study of the excited states of the one-dimensional Hubbard model \cite{jeckelmann1999}, where the evolution of the optical spectra was studied and the optical conductivity tuned by varying the ratio of $U$ to $t$. In the limit of small Mott gaps ($U \ll t$), a holon-anti-holon field theory was introduced to describe excitons, whereas for large Mott gaps ($U \gg t$), double occupancy and hole states were used (corresponding to the ionic states found in this paper).   By comparing our results for finite arrays obtained in this paper with those from an effective Hubbard model, one could determine the range of donor separations where the effective model is valid, as well as the best-fitting values of the model parameters. 

We have also compared the triplet-singlet (and quartet-doublet) energy splittings between a set of appropriate excited states for DA2 (DA3). From our calculations we can see that the optimal donor distance for optically operating a multi-qubit quantum gate is between 10 and 20 nm, where the ground-state exchange is below 0.02\,meV but the excited-state exchange is still considerable, as shown in Fig.\ref{fig:1} and Fig.\ref{fig:4}. A typical excited-state exchange interaction is $\approx 1 $ meV at $d\approx10\,\mathrm{nm}$, leading to a quantum gate operation time of $\approx 10$ ps, which is much shorter than the typical excited-state relaxation time ($\approx 200$ ps) of silicon donors. This result supports the realisation of the so-called `control-qubit' scheme \cite{sfg}.

Our results not only address the importance of correlations between donor electrons by using state-of-the-art first-principles tools, but also suggest a trend for the excited states and excitation energies as the number of donors increases, leading to an understanding of the electronic structure of periodic donor arrays. 
If further combined with CCC and MVE, this type of calculations could provide a more complete picture of the excited states of donor clusters in silicon. These two effects will make the donor orbitals more confined, so we expect the exchange interaction between the ground-state donor and the excited one would be reduced, lowering the optimal distance for quantum computing in the real silicon lattice and also leading to greater sensitivity to donor placement \cite{koiller2002}. The donor-array axis direction in silicon is also expected to have a significant effect on exchange interactions (as previously found  in the ground-state exchange \cite{koiller2002}); this will be investigated in a future publication. Moreover, these calculations could be useful to assess arrays of impurities in other host materials such as gallium arsenide (GaAs) and germanium, by adjusting the effective-mass parameters.

\begin{acknowledgments}
We wish to acknowledge the support of the UK Research Councils Programme under grant EP/M009564/1. We thank Eran Ginossar, Gabriel Aeppli, Neil Curson, Taylor Stock, Alex K\"{o}lker and Guy Matmon for helpful and inspiring discussions.
\end{acknowledgments}

%\newpage

\end{document}